\documentclass{SSA-SRL}

\citethisauthor{H.~Li et al.}
\doi{00.0000/000000000}
\recdate{00 Month 0000}

\graphicspath{{./Figures/}{./Logos/}}

\begin{document}

\title{Bridge Vibrations as Persistent Seismic Sources for Time-lapse Imaging of River Corridors with Distributed Acoustic Sensing}

\author[*1\orc{0000-0002-3230-8339}]{Haipeng Li}
\author[2\orc{0000-0001-5559-1627}]{Jingxiao Liu}
\author[3\orc{0000-0001-6060-7128}]{Doyun Hwang}
\author[3\orc{0000-0002-7998-3657}]{Hae Young Noh}
\author[1\orc{0000-0002-0961-2287}]{Biondo Biondi}

\affil[1]{Department of Geophysics, Stanford University, Stanford, California, U.S.A.}{}
\affil[2]{School of Civil and Environmental Engineering, Nanyang Technological University, Singapore.}{}
\affil[3]{Department of Civil and Environmental Engineering, Stanford University, Stanford, California, U.S.A.}{}

\corau{*Corresponding author: Haipeng Li, haipeng@sep.stanford.edu}

\begin{abstract}
Fluvial corridors mediate dynamic river--groundwater exchanges and support major transportation infrastructure, yet continuous subsurface monitoring remains limited by sparse point sensors and infrequent time-lapse surveys. Addressing this observational gap is critical during extreme flood events, when transient saturation, pore-pressure buildup, and structural scour drive complex hydro-mechanical changes that threaten both riparian slope stability and bridge foundations. Here, we demonstrate that traffic-excited highway bridges act as persistent, spatially fixed seismic sources; their vibrations, recorded on roadside telecommunication fiber with Distributed Acoustic Sensing (DAS), enable time-lapse imaging of river corridors. We evaluate this approach using dark fiber recordings along South Korea's Gyeongbu Expressway during a record monsoon flood in July 2025, targeting a $430\text{~m}$ multi-span bridge crossing the Miho River. Vehicle impacts excite flexural resonances of the bridge deck, generating a discrete spectral comb of fundamental-mode Rayleigh waves in the surrounding ground; these $3\text{--}6\text{~Hz}$ waves propagate beyond $400\text{~m}$ across both riverbanks to image the alluvial section down to bedrock. Because bridge-structure geometry constrains the source location and dominant modal frequencies, virtual source gathers achieve rapid convergence within $5\text{~min}$ of ballistic wave stacking with high waveform repeatability over the recording period. Relative seismic velocity ($\mathrm{d}v/v$) monitoring resolves velocity drops up to $1.0\%$ during flood passage associated with heavy precipitation and river stage rise, consistent with hydrological softening of the riparian sediments. These results demonstrate that traffic-excited bridges provide persistent, repeatable seismic illumination, turning roadside telecommunication cables into dense arrays for both hydrogeophysical and civil infrastructure applications.
\end{abstract}

\maketitle

\section{Introduction}

Fluvial corridors support major transportation infrastructure while mediating hydrogeological exchange between active river channels and unconfined alluvial aquifers \citep{harvey2015rivercorridor}. Within these corridors, deep foundation elements extend tens of meters into alluvial sediments that face structural degradation from scour and subsurface erosion \citep{wardhana2003bridgefailures,prendergast2013pilescourfrequency}. These geotechnical structures share space with dynamic hydrologic processes that govern bank stability and regional groundwater storage as saturation and pore pressure fluctuate \citep{rinaldi1999streambankstability,sophocleous2002gwswinteractions}. Extreme flood events amplify these combined risks by simultaneously accelerating scour and elevating pore pressure. Facing these compounding hazards, standard monitoring frameworks rely on sparse point sensors, isolated observation wells, and periodic manual surveys \citep{prendergast2014scourreview}. Overcoming these operational gaps requires continuous, spatially dense subsurface measurements to evaluate river--groundwater exchange, track bank stability, and safeguard bridge foundations during extreme hydrological forcing.

Hydrogeophysical methods offer a non-invasive framework for characterizing subsurface properties and tracking hydrologic processes over time \citep{binley2010hydrogeophysics}. Active-source seismic surveys constrain shear-wave velocity ($V_S$) structures across geotechnical and hydrologic depths \citep{xia1999estimation,pasquet2015watertable}, yet standard acquisition remains costly, labor-intensive, and restricted during severe weather events \citep{holbrook2026active}. Passive seismic methods overcome active-source limitations by utilizing ambient vibrations for continuous wavefield illumination \citep{sensschonfelder2006passive}. These passive ambient field recordings yield relative seismic velocity variations ($\mathrm{d}v/v$) that quantitatively track pore pressure \citep{fokker2023porepressure}, water-table elevation \citep{clements2018tracking,lecocq2017groundwater}, and soil saturation \citep{illien2021subsurface}, increasingly across aquifer scales \citep{rodrigueztribaldos2021aquifer,shen2025waterdynamics}. Deploying these passive seismic approaches along transportation corridors, however, is limited by the high cost of installing and maintaining dense geophone arrays. Distributed Acoustic Sensing (DAS) solves this challenge by converting pre-existing telecommunication fiber into dense seismic receiver arrays spanning tens of kilometers with meter-scale channel spacing \citetext{\citealp{dou2017distributed, zhan2020das,lindsey2021fiber}; \citealp{yuan2020nearsurface}; \citealp{liu2025urbandasreview, li2025nearsurface}}.

Despite dense sensor coverage, passive seismic monitoring across river corridors remains constrained by depth sensitivity and source spatio-temporal stability. Subsurface processes within unconfined aquifers and bridge foundations typically extend down to tens of meters, requiring surface waves of sufficiently low frequency to reach these depths. Along highway corridors, however, this depth-sensitive frequency range falls into a spectral gap bounded by ocean microseisms below $1\text{~Hz}$ \citep{mcnamara2004ambient} and surface vehicular traffic typically above $6\text{~Hz}$ \citep{nakata2011shear,liu2025characterizing}, which illuminate either too deep or too shallow. In addition to spectral coverage, time-lapse monitoring requires high source repeatability, as non-stationary moving sources can introduce spurious artifacts that bias estimates of true medium changes \citep{zhan2013spurious, garambois2019ballistic}. Overcoming these non-stationarity effects requires repetitive, spatially localized sources, such as train passages or fixed industrial machinery \citep{dales2017persistent,mordret2020noise,sobolevskaia2024aquiferdas,li2025fiberoptic,brenguier2019train,pinzonrincon2025attenuation,higueret2025trains}. However, characterizing river crossings is complicated by quiescent ambient energy in the depth-sensitive band below $10\text{~Hz}$ and continuously varying road traffic, making reliable corridor monitoring dependent on a spatially fixed, persistent source operating within an appropriate frequency band.

In this study, we exploit the bridge structure itself as a persistent, spatially fixed seismic source recorded by roadside fiber-optic arrays. Ambient traffic crossing the deck provides broadband structural excitation \citep{yang2005vehiclebridge,liu2023telecomdasbridge,rodet2025darkfiberbridge}, driving resonant vibrations through piers and abutments that radiate seismic waves directly into surrounding alluvial sediments \citep{gueguen2000buildingradiation,wu2004groundvibration,takemiya2007viaductground,kouroussis2014railwayreview}. These deck vibrations concentrate energy at a few hertz set by span length and flexural stiffness \citep{liu2026bridgehealth}, filling the above-mentioned spectral illumination gap to sample shallow water tables and foundation depths down to tens of meters. Unlike non-stationary sources such as moving road traffic or passing trains \citep{brenguier2019train,pinzonrincon2021trains,mukherjee2026anthropogenic}, a bridge injects seismic energy through fixed substructures throughout its operational life, similar to wind turbines and other fixed structures \citep{dales2017persistent,neuffer2021windturbinewavefield,sheng2023repeatingsources,mangriotis2026windturbinereview}. These stationary components radiate a discrete modal comb whose resonant frequencies are primarily governed by structural properties. Traffic volume mainly modulates radiation amplitude, whereas the modal frequencies vary only at the percent level through added vehicle mass and thermal effects well documented in structural health monitoring \citep{peeters2001z24environmental,kaya2013temptrafficmodal,furtmuller2022massloading}; larger shifts may indicate structural changes, so the recorded spectrum provides a means to assess source variability. Traffic-driven bridge vibrations combined with fiber-optic sensing thus convert river crossings into permanent, spectrally observable seismic nodes for continuous subsurface monitoring.

Despite their prevalence along transportation corridors, the use of bridge structures as seismic sources for subsurface imaging and monitoring remains relatively unexplored. In this study, we demonstrate an approach leveraging bridge infrastructure as a seismic source using ten days of dark-fiber DAS recordings along South Korea's Gyeongbu Expressway, captured during a major monsoon flood event in July 2025. Focusing on a $430\text{~m}$ multi-span bridge crossing the Miho River, we analyze the bridge-radiated wavefield, evaluate phase stability across interferometry, and measure subsurface velocity changes ($\mathrm{d}v/v$) under extreme rainfall conditions. We then examine these velocity variations in relation to precipitation and river stage rise and discuss their possible connection to alluvial groundwater dynamics, establishing practical benchmarks for continuous hydrogeophysical imaging and monitoring across river corridors.

\section{Miho River Bridge and DAS Experiment}

\begin{figure*}[!t]
    \centering
    \includegraphics[width=\textwidth]{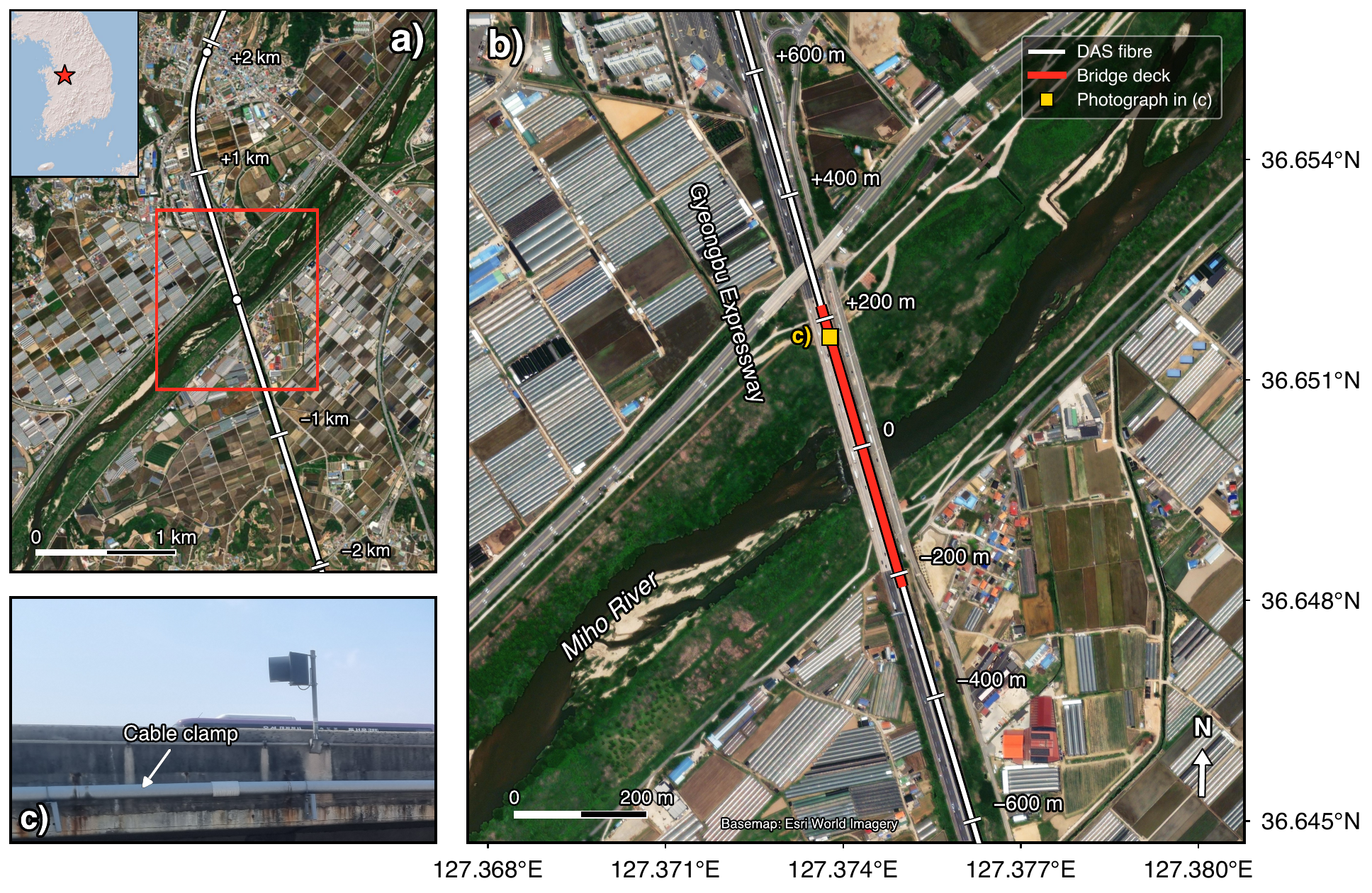}
    \caption{Field site and DAS array across the Miho River Bridge along the Gyeongbu Expressway, South Korea. (a) Regional corridor map illustrating the fiber-cable geometry (white line) and the target Miho River Bridge crossing; the inset locates the study area within South Korea. (b) Local array configuration at the Miho River Bridge crossing, where the roadside cable transitions onto the bridge deck to form a deck-coupled interval along the $430\text{~m}$ bridge (red line). (c) Field photograph showing the fiber-optic cable clamped to the exterior face of the bridge deck, corresponding to the yellow marker in b. Spatial positions indicate along-fiber distances relative to the center of the deck-coupled fiber interval at $0\text{~m}$.}
    \label{fig:site}
\end{figure*}

In this study, we analyze continuous DAS data acquired along the Gyeongbu Expressway near Cheongju, South Korea, using a dark telecommunication fiber-optic cable installed within a roadside conduit (Figures~\ref{fig:site}a and \ref{fig:site}b). An AP Sensing interrogator recorded dynamic strain rates at $1,000\text{~Hz}$ with a channel spacing of $1.225\text{~m}$ and a gauge length of $4.9\text{~m}$. To improve the signal-to-noise ratio, raw records were spatially averaged to an effective channel spacing of $4.9\text{~m}$ and decimated to $250\text{~Hz}$. Array channels were georeferenced using highway survey coordinates, locating the Miho River bridge crossing at $3.3\text{~km}$ along the fiber from the interrogator. This continuous nine-span concrete box-girder bridge was previously monitored for structural health with the same telecommunication fiber \citep{liu2026bridgehealth}. At this crossing, the cable exits the underground conduit and attaches directly to the exterior face of the bridge deck (Figure~\ref{fig:site}c). This deck-coupled interval spans the full $430\text{~m}$ bridge deck and is flanked on both sides by buried cable sections along the approach embankments. Because these subterranean approach sections run inline with energy radiating from the crossing, this geometry maximizes DAS axial strain sensitivity to outgoing Rayleigh waves. All array positions are referenced as along-fiber distances relative to the center of the bridge deck at $0\text{~m}$.

This study uses a dataset spanning 15--25 July 2025 to track subsurface responses through a major monsoon flood; two acquisition gaps interrupt the record, leaving nine days of data in total. All times are reported in Korea Standard Time (KST; UTC+9) unless otherwise noted. An initial 20-hour dry period from 15 to 16 July preceded intense precipitation concentrated between the afternoon of 16 July and the morning of 19 July. This heavy rainfall drove a sustained increase in river stage through 21 July, followed by a steady recession across the remaining four days of the experiment. This dynamic hydrological cycle provides a favorable opportunity to evaluate the spatio-temporal stability of bridge-radiated virtual source gathers and track time-lapse seismic velocity variations ($\mathrm{d}v/v$) during rapid alluvial saturation and drainage.

\section{Bridge-Radiated Seismic Wavefields}
\label{sec:wavefield}

\subsection{Wavefields in the DAS Records}

\begin{figure*}[!t]
    \centering
    \includegraphics[width=\textwidth]{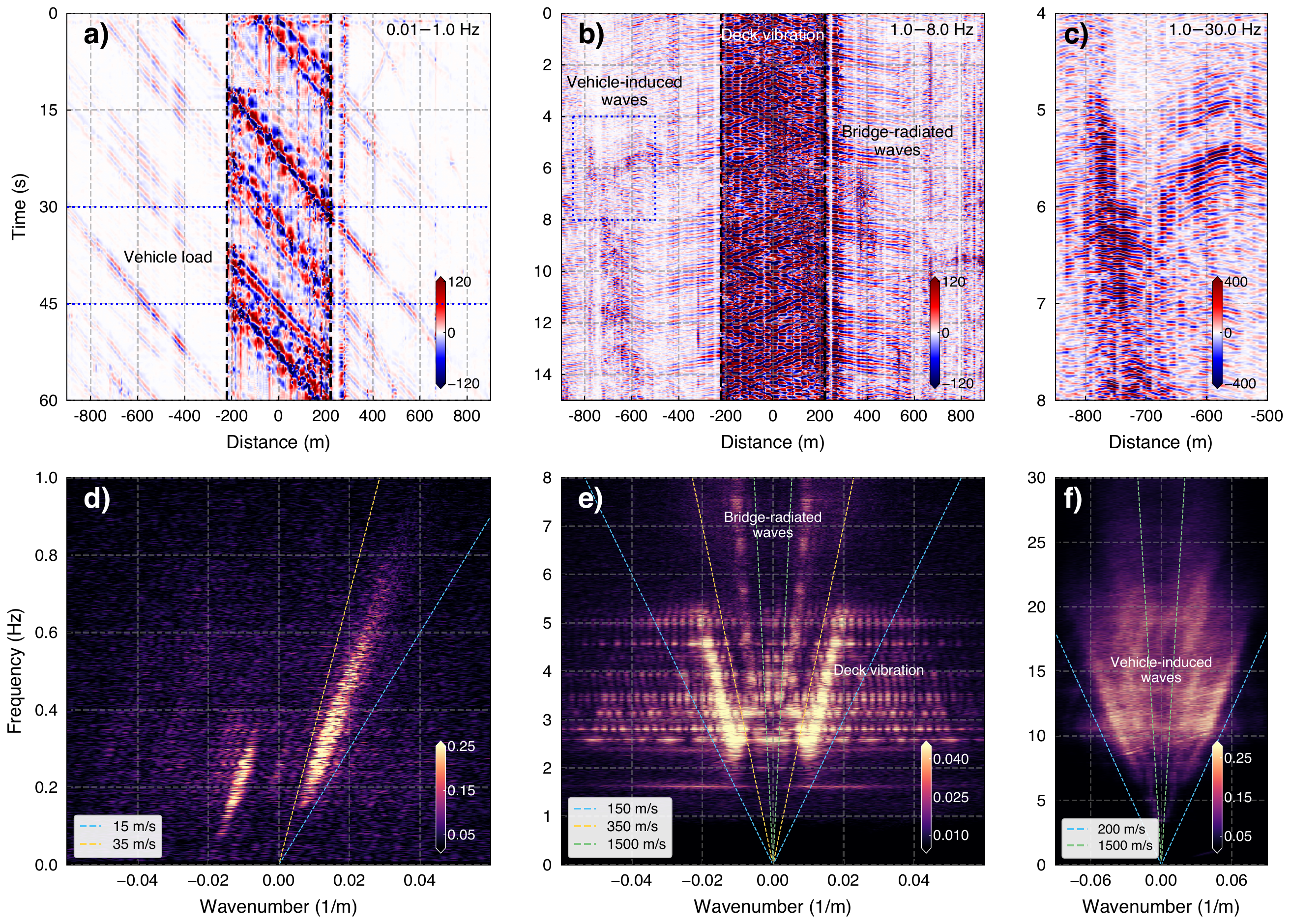}
    \caption{Strain-rate wavefields recorded across the Miho River Bridge crossing (09:05~KST, 16 July 2025). Distances are referenced to deck center ($0\text{~m}$), with dashed lines at deck ends. (a)~Quasi-static strain rate ($0.01\text{--}1\text{~Hz}$, $60\text{~s}$ window) showing vehicle load tracks. Blue dotted lines mark the $15\text{~s}$ window shown in panel~b. (b)~Bridge-generated seismic wavefield ($1\text{--}8\text{~Hz}$, $15\text{~s}$ window), showing continuous deck vibration radiating surface waves into both riverbanks alongside vehicle wavefields; the dotted box outlines the window in panel~c. (c)~Vehicle-induced wavefields on the far southern bank ($1\text{--}30\text{~Hz}$, $4\text{~s}$ window) generated by vehicles on the open roadway. (d--f)~$f$--$k$ spectra computed from $10\text{~min}$ of data beginning at the same time, over the channel ranges of panels~a and b (d, e) and the far-bank window of panel~c (f). In panel~e, the bright branches between the $150$ and $350\text{~m/s}$ lines are flexural waves guided along the deck girder. In panel~d, quasi-static energy aligns with vehicle speeds of $15\text{--}35\text{~m/s}$. In panel~e ($1\text{--}8\text{~Hz}$), deck vibration appears as horizontal bands at the modal frequencies, whose amplitude varies along wavenumber in lobes spaced by the inverse deck length, while energy between $350\text{~m/s}$ and $1500\text{~m/s}$ corresponds to bridge-radiated seismic waves. In panel~f (far-bank channels, up to $25\text{~Hz}$), the vehicle-induced field is broadband above roughly $6\text{~Hz}$ at apparent velocities of $200\text{--}1500\text{~m/s}$.}
    \label{fig:das_fk}
\end{figure*}

Continuous DAS strain-rate recordings across the fluvial corridor reveal distinct wavefield components partitioned by frequency band and spatial propagation character (Figure~\ref{fig:das_fk}). In the quasi-static band ($0.01\text{--}1\text{~Hz}$) recorded over a $60\text{~s}$ window, passing vehicle loads produce linear moveout trajectories whose slopes directly indicate vehicle speed (Figure~\ref{fig:das_fk}a). This quasi-static strain response amplifies prominently as wheel loads traverse the deck-coupled section and flex the bridge superstructure. Across the $1\text{--}8\text{~Hz}$ dynamic band shown over a $15\text{~s}$ window, dynamic traffic loads excite propagating elastic wavefields throughout the corridor (Figure~\ref{fig:das_fk}b). Across the river crossing, traffic excites high-amplitude flexural vibrations that travel along the continuous deck girder as guided structural waves \citep{liu2026bridgehealth}. At the southern and northern deck ends, the bridge abutments act as secondary seismic radiators, converting deck flexure into energetic surface waves that radiate outward into the surrounding alluvial sediment and remain detectable beyond $400\text{~m}$ offsets. In contrast, on open roadway segments away from the bridge structure, individual vehicle transits radiate localized, rapidly attenuating surface wave packets, as highlighted in the southern margin inset (dotted box in Figure~\ref{fig:das_fk}b, detailed in Figure~\ref{fig:das_fk}c) \citep{yuan2024vehicleinduced,liu2025characterizing}.

Frequency-wavenumber ($f$--$k$) spectra illustrate the propagating speeds and spatial characteristics of these wavefields (Figures~\ref{fig:das_fk}d--f). In the quasi-static band from $0.01\text{--}1\text{~Hz}$ (Figure~\ref{fig:das_fk}d), energy aligns along linear trajectories corresponding to vehicle speeds between $15\text{~m/s}$ and $35\text{~m/s}$ ($54\text{--}126\text{~km/h}$), dominated by a single travel direction because the fiber cable resides on one side of the highway. A second cluster with similar slope appears at negative wavenumbers, shifted by a constant offset of $\Delta k \approx 0.020\text{~m}^{-1}$. This feature corresponds to the first-order space harmonic ($k = f/v - 1/L$) from the span-by-span bridge response. For this continuous nine-span concrete box girder, the seven $L = 50\text{~m}$ interior spans yield $1/L = 0.020\text{~m}^{-1}$ \citep{liu2026bridgehealth}. This harmonic is present only on the bridge deck and not on the roadside fiber.

In the $1\text{--}8\text{~Hz}$ band (Figure~\ref{fig:das_fk}e), structural deck vibrations and propagating seismic waves appear as distinct spectral features. Horizontal spectral bands at the structural modal frequencies, clustered at several Hz, extend across wavenumbers beyond physical subsurface wave velocities and display interference sub-lobes spaced at the inverse of the deck length. These horizontal bands represent coherent deck vibrations. A slower branch confined to the deck interval, with phase velocities of $150\text{--}350\text{~m/s}$, corresponds to flexural bending waves guided along the girder \citep{liu2026bridgehealth}. Between apparent phase velocities of $350\text{~m/s}$ and $1500\text{~m/s}$, the same modal frequencies re-emerge as bridge-radiated seismic waves propagating through the alluvial deposits of the river corridor. The direct vehicle-induced seismic wavefield shows distinctly different spectral properties, as shown in Figure~\ref{fig:das_fk}f. Specifically, vehicle-induced energy is broadband rather than modal and concentrated above about $6\text{~Hz}$. These spectral differences reflect their underlying physical mechanisms, contrasting the narrowband resonant energy of a spatially fixed bridge structure with the high-frequency broadband excitations of moving axle impacts.

\subsection{Spectral Characterization}

\begin{figure*}[!t]
\centering
\includegraphics[width=\textwidth]{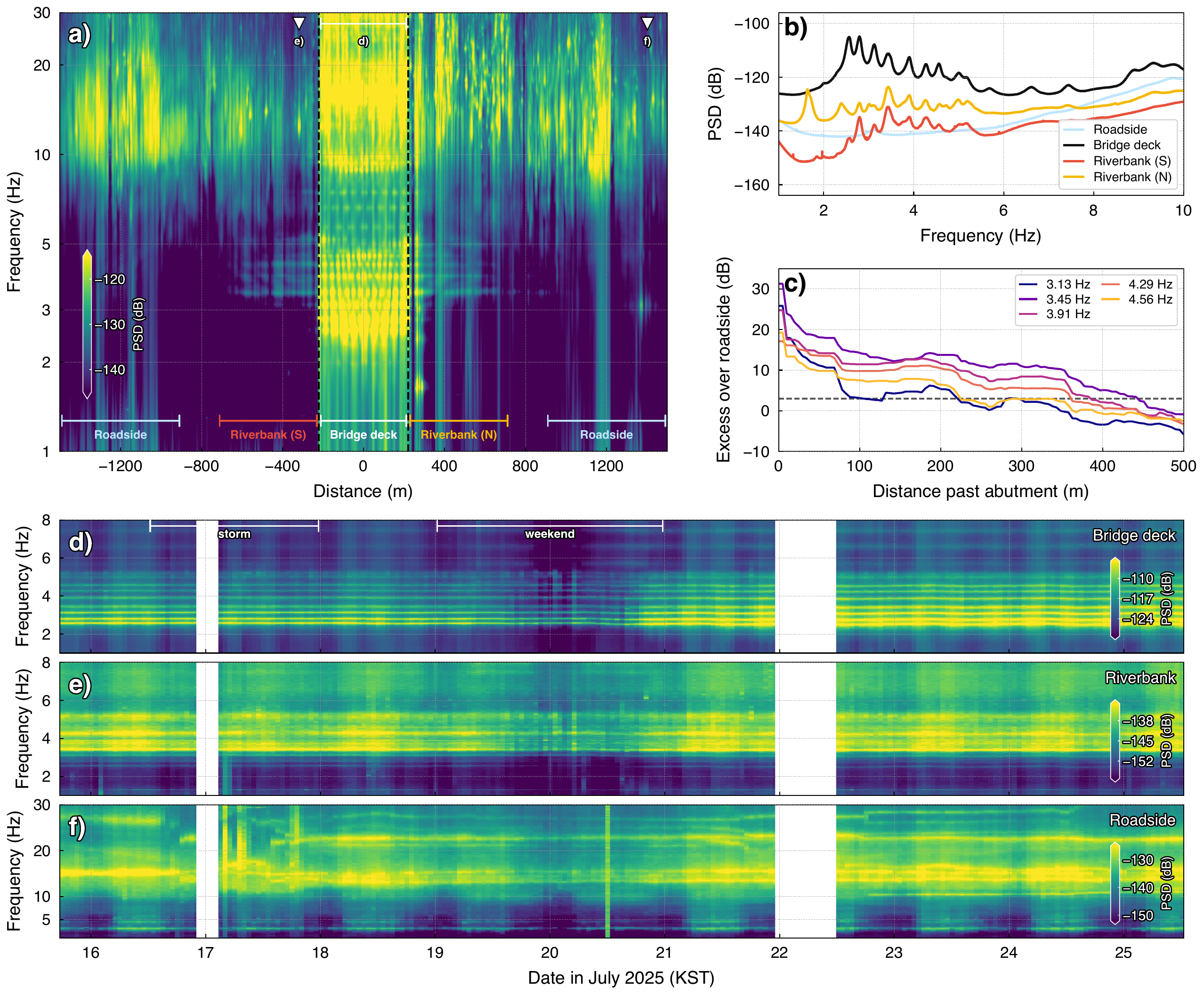}
\caption{Spectral characterization of the Miho River Bridge source derived from ten-day strain-rate PSD estimates calibrated in $\text{dB~re~}1\text{~strain}^2/\text{Hz}$. (a)~Mean PSD along the fiber array ($1\text{--}30\text{~Hz}$). Dashed vertical lines mark bridge deck boundaries, colored brackets define spatial zones, and triangles indicate channel locations for panels d--f. (b)~Zone-averaged spectra ($1\text{--}10\text{~Hz}$) showing the bridge deck in black, southern bank in red, northern bank in orange, and roadside reference in light blue. (c)~Spectral line power excess over the roadside reference plotted against absolute offset distance from the abutments. Dashed horizontal line marks the $3\text{~dB}$ detection threshold for radiated modal signals. (d--f)~Hourly spectrograms of the bridge deck ($1\text{--}8\text{~Hz}$), southern riverbank ($1\text{--}8\text{~Hz}$), and open roadside ($1\text{--}30\text{~Hz}$), with channel locations marked by triangles in panel a. The storm interval and the weekend are labeled above the spectrograms.}
\label{fig:source_spectral}
\end{figure*}

Calibrated strain-rate power spectral density (PSD) estimates show the spatial power distribution across the $3\text{~km}$ corridor (Figure~\ref{fig:source_spectral}a). Along the array, spectral power partitions into three distinct spatial domains. Across the bridge deck, mean power between $2\text{--}8\text{~Hz}$ reaches $-116.0\text{~dB}$, exceeding the $-136.5\text{~dB}$ roadside background by $20.5\text{~dB}$. This roadside reference still contains channels with elevated energy near a highway interchange ramp at $+1180\text{~m}$ and a local road crossing at $-1140\text{~m}$ (Figure~\ref{fig:source_spectral}a), which dominate the linear average. Consequently, the median across selected roadside channels is roughly $3\text{~dB}$ lower, so this roadside reference should be regarded as an overestimate. Beyond the crossing, the open roadside array records traffic noise above $6\text{~Hz}$ but contains relatively weak ambient energy below it, showing a frequency gap between $1\text{~Hz}$ and $6\text{~Hz}$. In adjacent riverbank channels, discrete spectral lines radiate outward from the bridge deck abutments past $400\text{~m}$ offsets. Power spikes in channels north of the bridge stem from strong energy emitted by a nearby overpass road (see Figure~\ref{fig:site}b), while other isolated channel-to-channel variations across the array reflect local differences in fiber coupling.

Figure~\ref{fig:source_spectral}b shows spectra averaged over three spatial zones, including the bridge deck, the riverbanks adjacent to the bridge, and the roadside reference. The bridge deck spectrum displays a modal comb at $2.6$, $2.8$, $3.1$, $3.4$, $3.9$, $4.3$, and $4.6\text{~Hz}$ rising well above roadside levels, whereas the roadside reference spectrum contains no distinct peak between $1\text{~Hz}$ and $12\text{~Hz}$. Along both riverbanks, this comb recurs at reduced amplitudes, showing wavefield transmission into the alluvial subsurface. A local road overpass north of the bridge injects vehicle energy into the alluvium through its piers and abutments. This secondary input accounts for the elevated energy on the north bank as well as the $1.6 \text{~Hz}$ spectral lines present only on that riverbank.

Propagation distance past the Miho River Bridge abutments displays strong frequency dependence (Figure~\ref{fig:source_spectral}c). At the bridge ends, a sharp $10\text{~dB}$ strain drop marks the transition from structural flexural strain to elastic ground strain, which offers potential opportunity for investigating soil-structure energy transfer mechanisms. Beyond the abutments, modal lines attenuate at frequency-dependent rates driven by intrinsic subsurface attenuation and geometric spreading, where higher-frequency modes generally decay faster. The most prominent radiating modes persist past $200\text{~m}$, reaching the $3\text{~dB}$ detection threshold (dashed horizontal line) between $360\text{~m}$ and $450\text{~m}$.

Ten-day spectrograms track the temporal pattern of these spectral features across those spatial zones (Figures~\ref{fig:source_spectral}d--f). On the bridge deck, the modal comb persists continuously (Figure~\ref{fig:source_spectral}d). Traffic volume modulates source amplitude by $2.7\text{~dB}$ diurnally and reduces weekend power, while the dominant resonant frequencies remain relatively stable. During peak monsoon flooding, total $2\text{--}8\text{~Hz}$ deck power dropped by only $1.2\text{~dB}$. This source stability contrasts sharply with highly dynamic environmental signals, such as discharge-dependent bedload noise \citep{roth2016bedload} and other hydrologically driven seismic sources \citep{dietze2022flood,roth2025riverfiber} that dominate many fluvial records. On the southern riverbank ($98\text{~m}$ offset), the radiating modal lines persist through heavy rainfall (Figure~\ref{fig:source_spectral}e). Along the roadside, $10\text{--}30\text{~Hz}$ traffic noise tracks daily vehicle activity but varies in both power and spectral content as traffic composition changes (Figure~\ref{fig:source_spectral}f). In particular, broadband bursts on 17~July coincide with the most intense rainfall and likely record rain-induced noise such as runoff through roadside drainage \citep{shen2023das}, while other increased spectral variations during the storm period suggest varied near-surface conditions. This nonstationarity makes raw traffic noise a less reliable illumination source. In contrast, the bridge acts as a natural mechanical band-pass filter, converting transient traffic impacts into sustained radiation within a persistent modal frequency band and providing a more repetitive seismic source for continuous subsurface monitoring.

\subsection{Propagation Physics and Mechanism}

The spectral comb originates from flexural vibrations of the multi-span bridge deck. For a uniform, simply supported Euler--Bernoulli beam, flexural natural frequencies scale as
\begin{equation}
f_n = \frac{n^2 \pi}{2 L^2} \sqrt{\frac{EI}{\rho A}}, \qquad n = 1, 2, \ldots,
\label{eq:beam}
\end{equation}
where $L$ represents span length, $EI$ is flexural rigidity, and $\rho A$ is mass per unit length \citep{chopra2017dynamics}. Instead of the wide quadratic frequency spacing expected for an isolated single span, the recorded spectrum contains closely spaced modes between $2.6\text{--}4.6\text{~Hz}$ with a mean interval of $0.3\text{~Hz}$. This modal clustering reflects elastic coupling across intermediate piers, which splits single-span frequencies into modal families that appear in $f\text{--}k$ space as horizontal ridges at fixed frequencies across a broad wavenumber range (Figure~\ref{fig:das_fk}e). Continuous traffic sustains these resonances because the average time between successive vehicles ($\sim 1\text{~s}$) is shorter than the structural damping time constant, preventing vibrations from decaying between passing vehicles. Operating as a lightly damped mechanical filter, the bridge converts broadband axle impacts into persistent harmonic radiation, allowing traffic volume to control signal amplitude while structural properties determine modal frequencies \citep{yang2005vehiclebridge,cantero2017evolution}. This mechanism is consistent with the relative stability of modal frequencies during traffic fluctuations, and distinguishes bridge modal radiation from transient expansion-joint impacts \citep{hou2022expansionjoint}.

Vibrational energy enters the subsurface through dynamic reaction forces at the bridge foundations. During deck oscillations, piers and abutments transmit dynamic vertical forces through their bearings into the underlying foundation soil, forming an array of point-force sources \citep{takemiya2007viaductground,kouroussis2014railwayreview}. Dynamic loading on sediments couples efficiently into seismic waves (Figure~\ref{fig:das_fk}e). High-frequency modes above $4.6\text{~Hz}$ dissipate rapidly near the abutments (Figure~\ref{fig:source_spectral}a) due to intrinsic material attenuation at short elastic wavelengths. 

Distinct spectral differences emerge across the fluvial corridor, characterized by a $5\text{--}10\text{~dB}$ power elevation on the northern riverbank and bank-specific modal peaks ($1.6\text{~Hz}$ on the north bank). We find they are primarily driven by secondary infrastructure, specifically an overpass where an intersecting roadway crosses above the highway fiber deployment, introducing localized traffic-induced energy. Other factors, such as spatial variations in cable-to-ground coupling, different abutment energy transmission, or lateral variations in alluvial subsurface properties, may also contribute to these spectral variations. Analyzing these spectral signatures offers a practical framework for evaluating soil-structure interaction, local site response, and fiber-coupling variability across the corridor.

\section{Bridge-Source Interferometry}
\label{sec:vsg}

\subsection{Virtual Shot Gathers and Dispersion}

To leverage the bridge as a persistent seismic source, we converted continuous DAS strain-rate recordings into VSGs through seismic interferometry \citep{campillo2003longrange,shapiro2004emergence}, following ambient noise interferometry on DAS arrays \citep{cheng2021submarine, yang2022subkilometer}. The detrended recordings were processed through a standardized ambient noise workflow \citep{bensen2007processing}. Records were band-passed between $1$ and $30\text{~Hz}$, resampled to $125\text{~Hz}$, segmented into $60$~s windows with $50\%$ overlap and one-bit normalized in the time domain, suppressing transient amplitude spikes while preserving phase information. Cross-correlations were computed up to a maximum lag of $6$~s, stacked over $5$-minute segments, and linearly stacked into $1$-hour bins.

\begin{figure*}[!t]
    \centering
    \includegraphics[width=\textwidth]{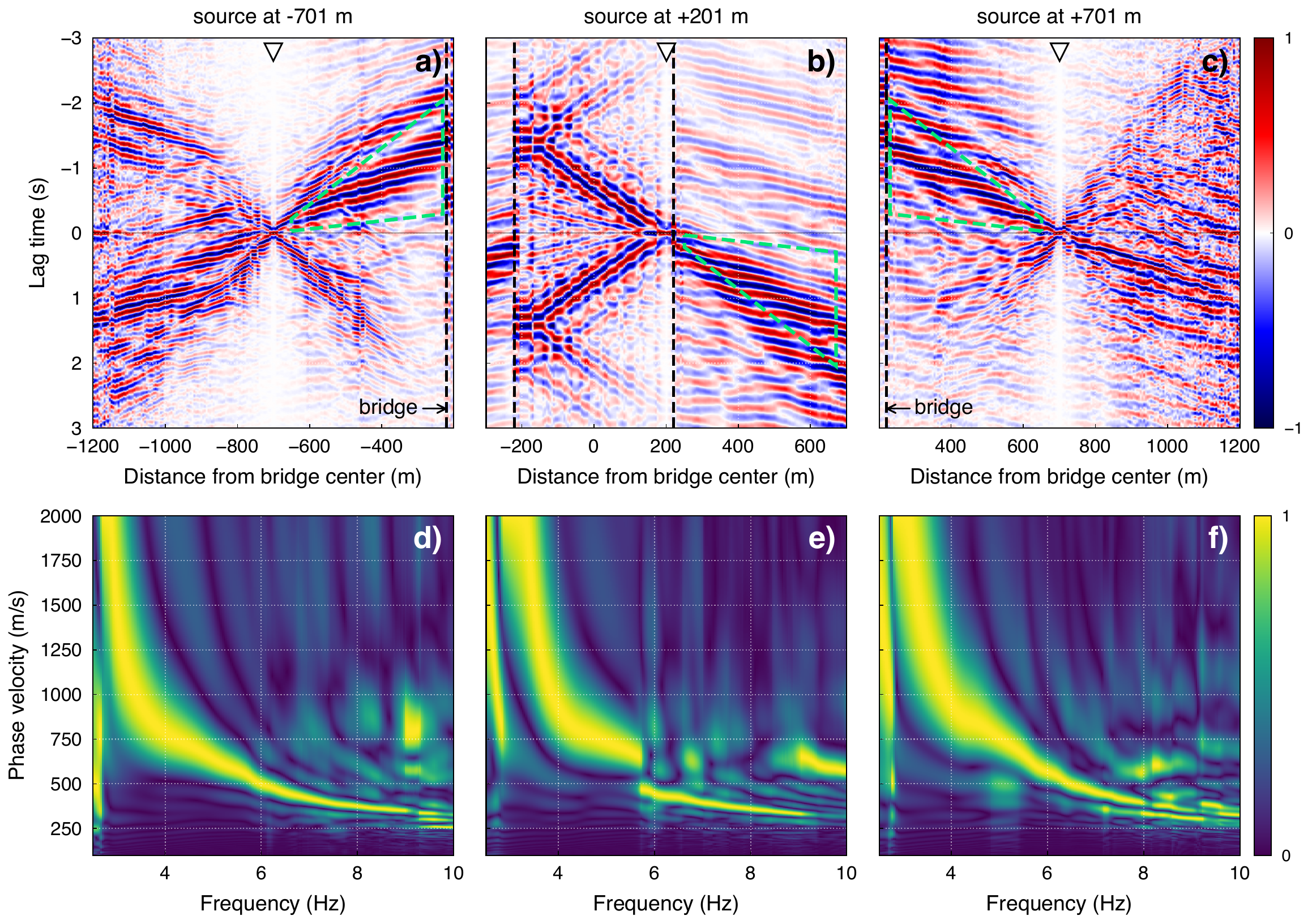}
    \caption{Virtual shot gathers (VSGs) and fundamental-mode Rayleigh wave dispersion across the Miho River crossing. (a--c)~All-record stacked VSGs ($2.5\text{--}10\text{~Hz}$) for virtual sources at $-701\text{~m}$, $+201\text{~m}$, and $+701\text{~m}$ relative to deck center (positive north, negative south). Dashed vertical lines denote deck boundaries, and triangles mark virtual source positions. Green dashed outlines indicate data selected for dispersion analysis. (d--f)~Phase-velocity spectra displaying a coherent fundamental-mode Rayleigh branch from $3\text{--}10\text{~Hz}$.}
    \label{fig:vsg}
\end{figure*}

To examine spatial wavefield behavior across the array, we selected three representative virtual sources on the bridge deck ($+201\text{~m}$) and on each fluvial margin ($-701\text{~m}$ and $+701\text{~m}$, located around $480\text{~m}$ beyond the southern and northern abutments, respectively). Figures~\ref{fig:vsg}a--c show the ten-day stacked VSGs for these locations, with dashed vertical lines marking the bridge deck boundaries. Because cross-correlations are driven by a spatially fixed structural source, each gather exhibits spatial asymmetry relative to its virtual source position. For the southern margin source at $-701\text{~m}$ (Figure~\ref{fig:vsg}a), the bridge-radiated wavefield forms an acausal branch propagating northward toward the bridge and a causal branch propagating southward away from it. The northern margin source at $+701\text{~m}$ displays an equivalent counterpart (Figure~\ref{fig:vsg}c). For the deck source at $+201\text{~m}$ (Figure~\ref{fig:vsg}b), a causal branch radiates northward into the riverbank, whereas flexural deck vibrations appear as a symmetric wavefield across causal and acausal lags and confined to the bridge deck, consistent with standing-wave structural dynamics \citep{liu2026bridgehealth}. Secondary moveout branches originate from vehicle traffic along the adjacent roadway. Because the optical fiber is deployed along one side of the highway, vehicles in the lanes nearest the fiber generate stronger traffic-induced moveout branches than those in the opposite-direction lanes.

Fundamental-mode Rayleigh wave phase velocity dispersion was extracted from the VSGs across channel offsets of $50\text{--}470\text{~m}$ using the phase-shift method (Figures~\ref{fig:vsg}d--f). The resulting phase-velocity spectra display a coherent, continuous fundamental-mode dispersion branch spanning $3\text{--}10\text{~Hz}$. Within this branch, bridge modal radiation supplies the energy below about $6\text{~Hz}$, whereas the higher frequencies are illuminated by vehicle-induced surface waves (Figure~\ref{fig:das_fk}f). The three virtual sources yield similar dispersion behavior (Figures~\ref{fig:vsg}d--f). Two physical constraints bound dispersion recovery. At low frequencies ($< 3\text{~Hz}$), Rayleigh wavelengths approach the receiver array aperture, broadening the dispersion ridge, whereas at higher frequencies ($> 10\text{~Hz}$), modal source energy is absent in the bridge spectrum and ambient ground signals attenuate rapidly within the shallow near surface.

\subsection{Stacking Convergence and Temporal Stability}

\begin{figure*}[!t]
    \centering
    \includegraphics[width=\textwidth]{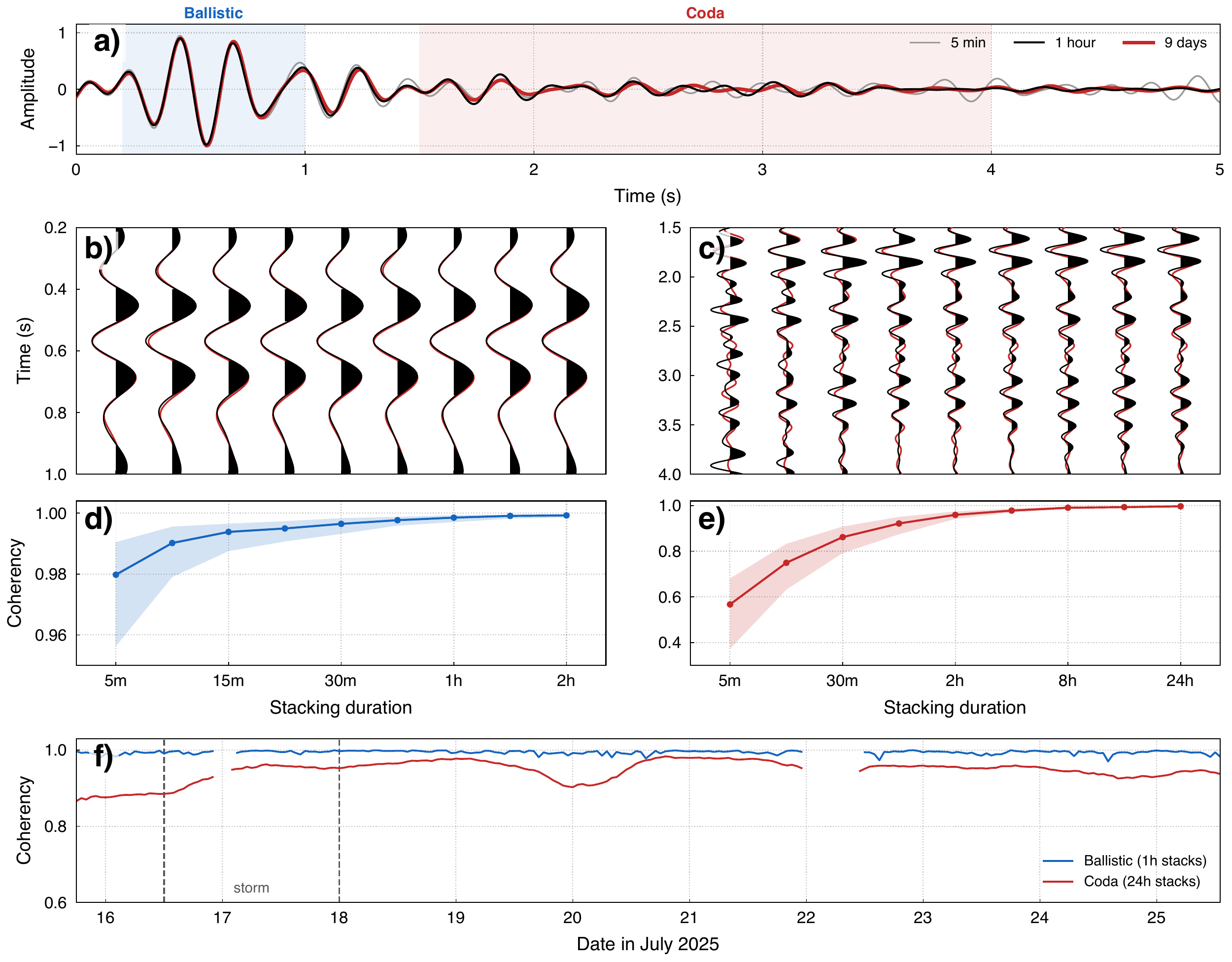}
    \caption{Stacking convergence and temporal stability of VSGs for the $+284\text{~m}$ virtual source and a receiver at $+485\text{~m}$, $201\text{~m}$ north of the source ($3\text{--}6\text{~Hz}$ bandpass). (a)~Cross-correlation functions stacked over $5\text{~min}$, $1\text{~h}$, and the full record (9 effective days of data for 10 days' recordings); shading highlights the ballistic ($0.2\text{--}1.0\text{~s}$) and coda ($1.5\text{--}4.0\text{~s}$) windows. (b, c)~Cumulative stacks for (b) the ballistic arrivals ($5\text{~min}$ to $2\text{~h}$) and (c) the coda arrivals ($5\text{~min}$ to $24\text{~h}$), overlaid on the $9\text{-day}$ reference in red (each panel scaled to its own amplitude). (d, e)~Stacking coherence relative to the $9\text{-day}$ reference for (d) ballistic and (e) coda windows. (f)~Time-series coherence relative to the reference evaluated at window-appropriate stacking durations, with $1\text{-hour}$ stacks for the ballistic window and $24\text{-hour}$ stacks for the coda. Dashed lines bound the period of the heavy rainfall.}
    \label{fig:vsg_stability}
\end{figure*}

To evaluate the stacking convergence and temporal stability of the bridge-radiated wavefield, we quantify cross-correlation coherence using a virtual source at $+284\text{~m}$ and a receiver at $+485\text{~m}$, $201\text{~m}$ north of the source ($3\text{--}6\text{~Hz}$ bandpass, the bridge-radiated band), as shown in Figure~\ref{fig:vsg_stability}. We compute virtual source gathers (VSGs) in 5-minute intervals and measure cumulative stack coherence relative to a full reference stack, which spans 10 calendar days with 9 effective days of data. Both ballistic and coda time windows were analyzed, including a ballistic window ($0.2\text{--}1.0\text{~s}$) capturing the direct arrival and a coda window ($1.5\text{--}4.0\text{~s}$) beginning several wavelengths behind the direct wave (Figure~\ref{fig:vsg_stability}a). Because these two windows differ by an order of magnitude in amplitude, their stacking convergence rates diverge significantly (Figures~\ref{fig:vsg_stability}b--c). Quantitatively, the ballistic arrival achieves a coherence of $0.98$ relative to the reference after a single 5-minute stack, exceeds $0.99$ within 10 minutes, and closely matches the reference within one hour (Figures~\ref{fig:vsg_stability}b and \ref{fig:vsg_stability}d). Individual 5-minute correlations already achieve coherency at a median of $0.96$ in the ballistic window, demonstrating its highly repeatable character. In contrast, the scattered coda field converges more than an order of magnitude slower, with the coherence of a single 5-minute stack against the reference rising from $0.57$ to $0.95$ at 2 hours and $0.99$ at 8 hours (Figure~\ref{fig:vsg_stability}e).

When analyzed over their respective convergence timescales, both signals remain highly stable throughout the flood event (Figure~\ref{fig:vsg_stability}f). Hourly stacks of the ballistic window maintain a median coherence of $0.995$ (minimum $0.97$) relative to the 9-day reference, while daily stacks of the coda window achieve a median coherence of $0.95$ (minimum $0.87$). The coda coherence is lowest before the storm and decreases again around 20~July during the weekend traffic minimum, whereas the ballistic coherence shows no such variation. The ballistic wavefield, induced by a source geometry fixed in concrete, is reproduced within minutes and retains high coherence through the flood event, whereas the scattered coda field sampling the surrounding medium requires several hours of averaging to emerge. These results demonstrate rapid recovery of repeatable ballistic waveforms, but do not by themselves establish the precision of velocity estimates at 5-minute intervals. For velocity monitoring, we use 1-hour correlation stacks to improve the stability of small traveltime-shift measurements.

\section[Time-lapse Imaging of River Corridors]{Time-lapse Imaging of River Corridors}
\subsection{Subsurface Structure of the Fluvial Corridor}
\label{sec:imaging}

\begin{figure*}[!t]
    \centering
    \includegraphics[width=\textwidth]{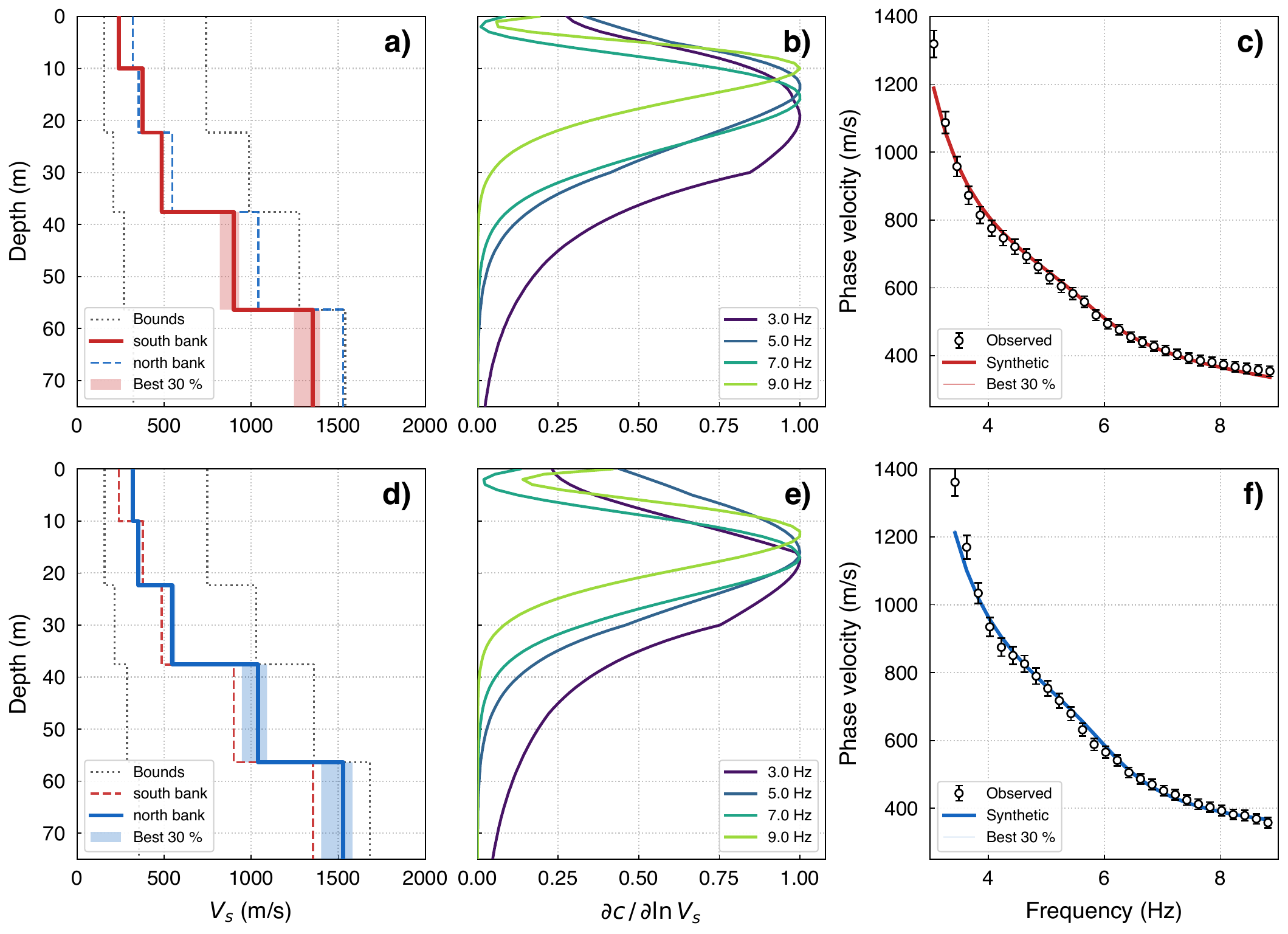}
    \caption{Shear-wave velocity structure across the fluvial corridor derived from dispersion inversion on both river margins. (a, d)~Inverted 1-D $V_S$ profiles for the southern (red) and northern (blue) riparian margins, displayed alongside the envelope of the top $30\%$ sampled models, the opposing margin profile (dashed line), and search bounds (dotted lines). (b, e)~Normalized fundamental-mode Rayleigh phase-velocity sensitivity kernels ($\partial c / \partial \ln V_S$) for representative frequencies between $3$ and $9\text{~Hz}$ computed from the fitted profiles, where $c$ denotes Rayleigh-wave phase velocity. (c, f)~Observed dispersion picks with error bars plotted against synthetic dispersion curves calculated from the best-fit models.}
    \label{fig:inversion}
\end{figure*}

To image the subsurface velocity structure across the fluvial corridor, fundamental-mode Rayleigh wave dispersion curves were inverted to retrieve 1-D shear-wave velocity ($V_S$) profiles for the southern and northern fluvial margins. We focus on the $3\text{--}10\text{~Hz}$ frequency range, where bridge modal radiation ($3\text{--}6\text{~Hz}$) and vehicle-induced surface waves ($6\text{--}10\text{~Hz}$) together yield a clearly developed fundamental-mode Rayleigh branch (Figures~\ref{fig:vsg}d, f). Resolving the lowest frequencies requires a wide receiver aperture because Rayleigh wavelengths at these frequencies exceed $400\text{~m}$, comparable to the $420\text{~m}$ array aperture. Consequently, the resulting dispersion measurement represents an aperture-averaged profile beneath the array. Spatial rolling of the virtual source along the array yields minimal independent structural information. Observed phase velocities decrease smoothly from around $1300\text{~m/s}$ at $3\text{~Hz}$ to $300\text{~m/s}$ at $10\text{~Hz}$ on both margins (Figures~\ref{fig:inversion}c, f). Dispersion inversion was performed using a particle swarm optimization algorithm, with forward dispersion curves calculated using fundamental-mode algorithms \citep{luu2023evodcinv, herrmann2013computer}. In the inversion, $V_S$ was allowed to vary between $0.40$ and $1.90$ times an initial reference model, with compressional velocity ($V_P$) and density coupled through empirical relations. Initial reference models were constructed by mapping phase-velocity picks to depth using $z = \lambda / 2.5$. Optimization was executed with $20$ particles over $200$ iterations and restarted $30$ times using independent random seeds. The best-fitting $30\%$ of all sampled models were retained to construct solution envelopes and characterize variability among the retained models (Figures~\ref{fig:inversion}a, d).

The inverted $V_S$ profiles closely match the observed dispersion data across both fluvial margins (Figures~\ref{fig:inversion}c, f), revealing a classic alluvial valley sequence on both sides of the active river channel. Low shear-wave velocities ($250\text{--}350\text{~m/s}$) in the upper $10\text{~m}$ correspond to unconsolidated overbank deposits, which stiffen progressively through compacting fluvial alluvium to about $900\text{--}1000\text{~m/s}$ at the base of the valley fill. This velocity-depth gradient reflects granular sediments under increasing effective overburden stress \citep{bachrach2000sands}. A sharp velocity impedance contrast at around $56\text{~m}$ depth, where $V_S$ increases to $1350\text{--}1550\text{~m/s}$, is interpreted as the transition to consolidated material beneath the alluvium. The northern margin is systematically faster than the southern margin, exhibiting a shear-wave velocity in the upper $30\text{~m}$ ($V_{S30}$) of $374\text{~m/s}$ compared to $332\text{~m/s}$ on the south. The retained model ensemble shows less spread in the shallow subsurface and greater variability below the interpreted bedrock interface. Normalized fundamental-mode Rayleigh wave sensitivity kernels calculated between $3$ and $9\text{~Hz}$ demonstrate peak depth sensitivity between $10$ and $20\text{~m}$, while retaining measurable sensitivity beyond $70\text{~m}$ depth at lower frequencies (Figures~\ref{fig:inversion}b, e). The bridge-radiated frequencies ($3\text{--}6\text{~Hz}$) carry the sensitivity to the deeper alluvium and the bedrock interface, whereas the vehicle-induced energy above $6\text{~Hz}$ mainly refines the uppermost ${\sim}15\text{~m}$. This sensitivity depth window encompasses the alluvial section, from the shallow unconfined aquifer down to the bedrock boundary.

Beyond structural imaging, these inverted velocity models provide valuable hydrogeological and geotechnical insights. The retrieved $V_S$ models and layering provide constraints on sediment stiffness and the thickness of the alluvial section. From a civil engineering standpoint, the $V_S$ profiles quantify subsurface soil stiffness and site classification, providing critical parameters for structural safety and dynamic soil-structure interaction assessments. The $400\text{--}900\text{~m/s}$ sedimentary package is relevant to foundation--soil interaction, although its relationship to pile embedment requires independent structural information. The shallow zone of maximum kernel sensitivity is also relevant to monitoring changes within the alluvial aquifer. Furthermore, the bedrock interface depth can provide foundational structural inputs for integrated surface water--groundwater models of the Miho watershed \citep{chung2015mihocheon}.

\subsection{Fluvial Corridor Monitoring During Heavy Rainfall}
\label{sec:monitoring}

\begin{figure*}[!t]
    \centering
    \includegraphics[width=\textwidth]{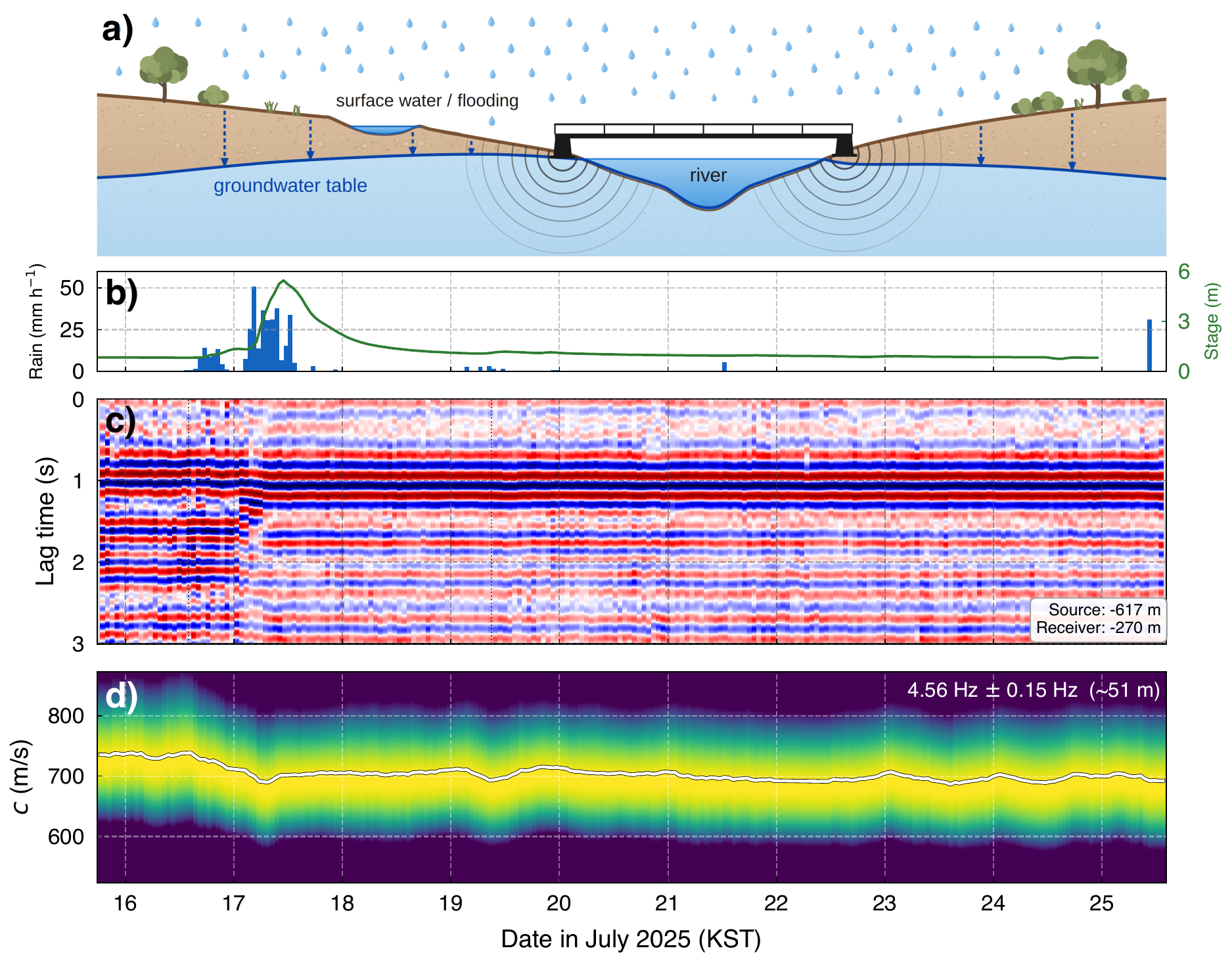}
    \caption{Time-lapse seismic data through the July 2025 rainfall event. (a)~Conceptual setting showing traffic-driven bridge deck vibrations radiating through pier and abutment foundations into riverbank alluvium during rainfall infiltration and water-table rise. (b)~Hourly precipitation recorded at the local meteorological station. (c)~Stacked hourly gather trace for a southern riverbank virtual source pair ($617\text{~m}$ south of deck center, $348\text{~m}$ offset). (d)~Phase-velocity tracking of the $4.6\text{~Hz}$ modal dispersion ridge, showing a velocity drop after the heavy rain.}
    \label{fig:monitor}
\end{figure*}

Relative seismic velocity variations ($\mathrm{d}v/v$) in unconfined alluvial aquifers can respond to changes in the degree of saturation, pore fluid pressure, and effective stress \citep{clements2018tracking,gaubertbastide2022watertable,li2025fiberoptic}. A major monsoon storm occurring in the middle of the ten-day recording period provided a natural hydrologic forcing across the fluvial corridor. Under such heavy rainfall conditions when subsurface structural changes occur most rapidly, repeated active-source time-lapse surveys would be logistically challenging or impractical. Continuous, high-intensity precipitation began at 14:00\text{~KST} on 16\text{~July} and lasted until 09:00\text{~KST} on 19\text{~July 2025}, peaking at $51\text{~mm/h}$ at 04:00\text{~KST} on 17\text{~July} (Figure~\ref{fig:monitor}b). Because bridge foundations transmit persistent modal vibrations directly into underlying foundation sediments, this transient rainfall event enabled continuous passive seismic monitoring of dynamic river--groundwater interactions across the river corridor (Figure~\ref{fig:monitor}a).

Time-lapse velocity variations are directly observable within individual VSGs. Figure~\ref{fig:monitor}c illustrates a representative virtual source pair on the southern riverbank ($617\text{~m}$ south of deck center, $348\text{~m}$ receiver offset) exhibiting high hourly trace coherence across the deployment. Ballistic Rayleigh waves retain relatively stable phase, whereas later arrivals show progressively increasing delays from the night of 16--17\text{~July} onward, which is consistent with a reduction in medium velocity, although source-related waveform changes cannot be excluded \citep{zhan2013spurious}. Tracking phase velocities along the $4.6\text{~Hz}$ modal dispersion ridge (sampling a depth proxy $\lambda/3 \approx 50\text{~m}$) resolves a $7.1\%$ velocity drop from $730\text{~m/s}$ during the dry baseline down to $678\text{~m/s}$ following peak rainfall (Figure~\ref{fig:monitor}d).

\begin{figure*}[!t]
    \centering
    \includegraphics[width=\textwidth]{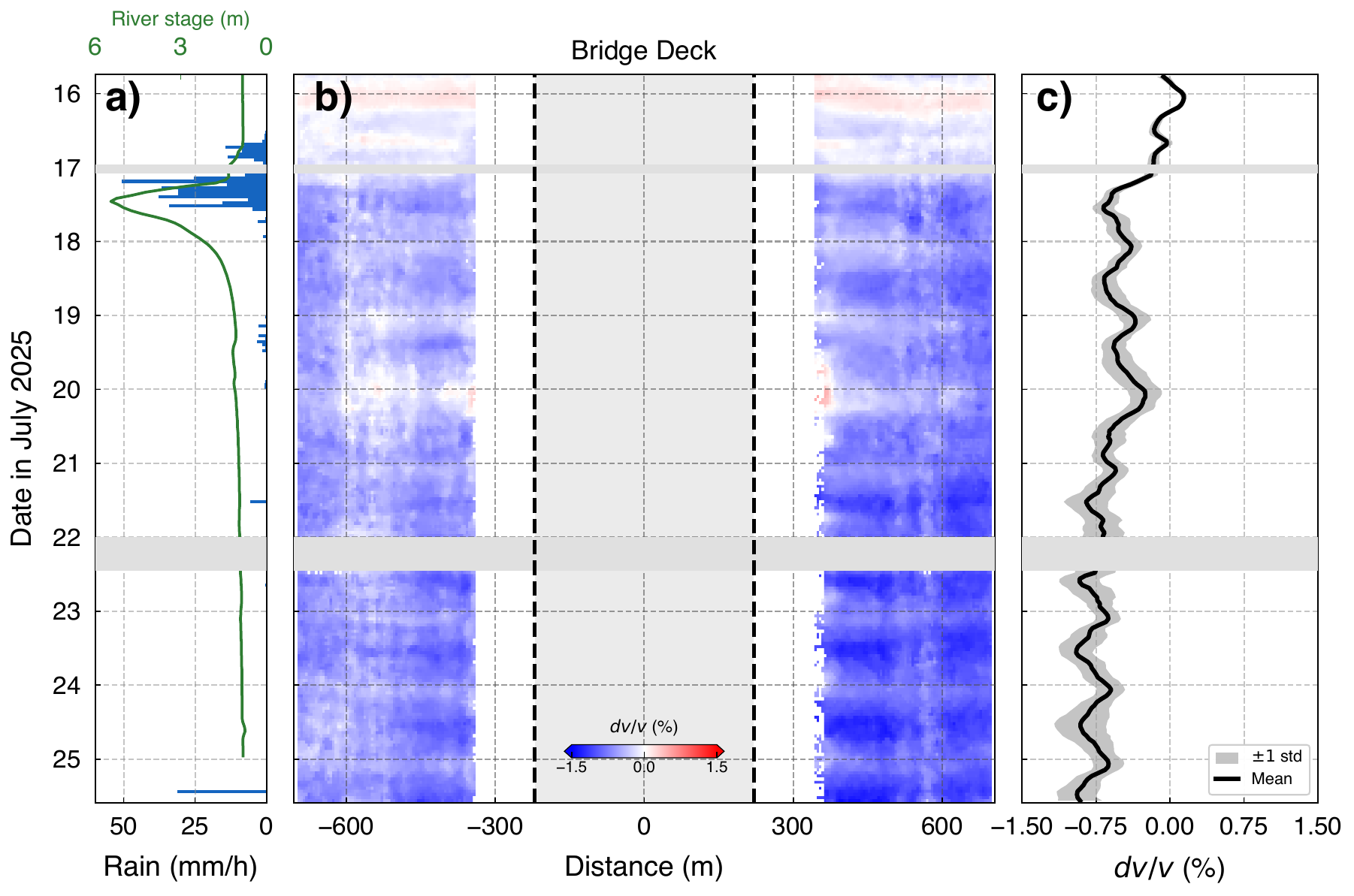}
    \caption{Spatiotemporal velocity monitoring across the fluvial corridor. (a) Hourly rainfall (bars) and Miho River stage (green). (b) Midpoint-stacked bridge-source $\mathrm{d}v/v$ section computed from all virtual source positions ($3$--$6\text{~Hz}$), showing coherent velocity changes on both riverbanks relative to the 20-hour pre-storm baseline; dashed lines mark the bridge deck edges. (c) Spatially averaged mean $\mathrm{d}v/v$ after excluding the upper and lower $7.5\%$ of spatial values at each time, with its $\pm1\sigma$ spread. Gray horizontal bands indicate periods without DAS acquisition.}
    \label{fig:dvv}
\end{figure*}

To extend time-lapse monitoring along the river corridor, $\mathrm{d}v/v$ analysis was performed at all virtual source positions spanning the ${\sim}3\text{~km}$ array, spaced at the channel interval of $4.9\text{~m}$. Processing was restricted to the $3\text{--}6\text{~Hz}$ frequency band, where the bridge-radiated wavefield is strongest and the fundamental-mode Rayleigh branch is clearly resolved, excluding the vehicle-dominated energy above $6\text{~Hz}$. To favor bridge-radiated energy over direct traffic noise, cross-correlations were selected according to the expected outward propagation from the bridge. Each monitor gather represented a $1\text{-hour}$ correlation stack (i.e., hourly resolution). A pre-storm average of $20$ hourly gathers between 18:00~KST on 15~July and 14:00~KST on 16~July formed the reference. Relative traveltime delays were measured using the moving-window cross-spectral (MWCS) method \citep{poupinet1984monitoring,clarke2011assessment} and converted to velocity variations via $\mathrm{d}v/v = -\mathrm{d}t/t$. The bridge-source product retained same-bank source--receiver paths directed toward the bridge and the outward-propagating correlation branch. Pairwise $\mathrm{d}v/v$ estimates were assigned to their source--receiver midpoints and stacked within midpoint bins. This procedure yielded a spatially continuous $\mathrm{d}v/v$ profile along the river corridor (Figure~\ref{fig:dvv}b).

Figure~\ref{fig:dvv}b shows the midpoint-stacked $\mathrm{d}v/v$ section along the corridor, where both riverbanks exhibit coherent velocity variations. Seismic velocities decrease upon storm arrival on 16--17\text{~July}, remain reduced during precipitation, rebound briefly around 20~July during the weekend traffic minimum (Figure~\ref{fig:source_spectral}d), and subsequently decline to a sustained minimum during late recession, reaching a maximum relative velocity drop of $1.0\%$ (Figure~\ref{fig:dvv}b). Superimposed on this response, hourly $\mathrm{d}v/v$ carries diurnal and weekly modulations that follow traffic volume (Figure~\ref{fig:source_spectral}d). Although the dominant modal frequencies remain relatively stable, changes in traffic may alter the balance between bridge radiation and other traffic-induced energy in the correlations, potentially introducing apparent velocity variations alongside medium changes \citep{zhan2013spurious}. The spatial section in Figure~\ref{fig:dvv}b shows lateral variability, although bins closest to the deck are sparsely resolved because few eligible source--receiver pairs remain after quality control. These $\mathrm{d}v/v$ observations are consistent with mechanical sediment softening operating over different timescales. Infiltration and water-table elevation can reduce vadose-zone matric suction and increase pore pressure, lowering effective stress and shear modulus ($\mu$) in unconsolidated alluvium \citep{bachrach2000sands,illien2021subsurface,mao2025stormrecovery, li2025fiberoptic}, while added pore water increases bulk density ($\rho$) to further lower shear-wave velocity ($V_S = \sqrt{\mu/\rho}$). The fast component tracks surface forcing, as velocities decrease with rainfall and the 17\text{~July} stage crest (Figure~\ref{fig:dvv}a). Conversely, the slow component lags by several days potentially reflecting storm recharge through the vadose zone and a delayed pore-pressure response after surface forcing subsides. Such a lag may depend on vadose-zone thickness and hydraulic properties, although lateral recharge from the river may also contribute. Spatial variations in vadose-zone thickness and initial saturation could contribute to differences between near-river and inland responses. The depth sensitivity of the $3\text{--}6\text{~Hz}$ signal makes delayed subsurface wetting and pore-pressure changes plausible contributors to the persistent velocity reduction, but does not uniquely identify groundwater-level changes or their depth. A quantitative interpretation linking $\mathrm{d}v/v$ amplitude and timing to recharge flux, vadose-zone thickness, and hydraulic properties is beyond the scope of this study.

\section{Discussion}

The traffic-driven wavefield along the fluvial corridor comprises two distinct components, namely direct vehicle-generated waves and bridge-radiated waves. Both propagate through the shallow near-surface as seismic waves (Figure~\ref{fig:das_fk}), yet their source mechanics differ fundamentally. Individual vehicles act as transient, moving, broadband sources whose energy distribution depends on speed, weight, and pavement roughness \citep{liu2025characterizing}. Reconstructing stable empirical Green's functions from direct vehicle noise usually requires extensive temporal averaging over repeated passages to overcome source non-stationarity \citep{nakata2011shear,brenguier2019train,yuan2024vehicleinduced}. Conversely, the bridge deck functions as a spatially fixed, lightly damped mechanical filter that converts stochastic axle impacts into persistent harmonic radiation governed by the flexural modes of the structure \citep{yang2005vehiclebridge}. The bridge constrains the dominant modal frequencies and the locations of energy transmission into the subsurface. Although traffic excitation remains stochastic, these structural constraints promote repeatable seismic illumination. This repeatability does not require a temporally invariant source, as traffic and environmental conditions may still alter the relative contributions of structural modes and radiating substructures. Because the dominant radiated modal band occupies the corridor's $3\text{--}6\text{~Hz}$ background noise gap (Figure~\ref{fig:source_spectral}b), the bridge injects high-amplitude, narrowband energy into an otherwise quiescent spectrum. Within its $\sim 450\text{~m}$ illumination aperture, five-minute correlation gathers already yield reference data with high coherence ($r = 0.96$), demonstrating rapid recovery of repeatable ballistic waveforms (Figure~\ref{fig:vsg_stability}).

Dark fiber deployment near infrastructure crossings enables simultaneous recording of structural dynamics and subsurface velocity variations. Structural health monitoring evaluates shifts in deck resonant frequencies \citep{brownjohn2007shmcivil,liu2023telecomdasbridge}, whereas near-surface geophysics extracts relative seismic velocity changes in the subsurface medium \citep{mordret2020noise}. Our study illustrates the potential dual utility of this approach, as the monsoon storm coincided with a peak $\mathrm{d}v/v$ drop of $1.0\%$ across the riparian aquifer (Figure~\ref{fig:dvv}b), whereas the bridge maintained stable resonant frequencies (Figure~\ref{fig:source_spectral}d). Although environmental factors such as thermal expansion or hydrodynamic added mass from elevated river stage can alter deck resonance \citep{peeters2001z24environmental,hernandez2023waterlevelmodal}, observing both the source and the propagated wavefield provides a basis for evaluating structural and subsurface changes, provided that their contributions to the recorded signals can be distinguished.

The spectral and spatial properties of bridge-radiated wavefields follow predictable structural scaling laws. For a uniform, simply supported bridge span of length $L$, flexural natural frequencies scale as $f_n \propto L^{-2} \sqrt{EI / \rho A}$. Individual span length, rather than total crossing length, determines the fundamental modal frequencies. In turn, the radiated frequency $f$ and local phase velocity $v_c$ establish the surface-wave wavelength $\lambda = v_c / f$, which defines the depth sensitivity range. In our study, the spatial aperture of this passive monitoring framework is bounded to approximately $450\text{~m}$ from the abutments by two distinct physical mechanisms, namely intrinsic material attenuation in shallow sediments at high frequencies and low-frequency radiation inefficiency governed by structural dynamics. By extension, this framework can be adapted to transform other civil infrastructure, such as railway viaducts, dams, or wind turbines, into persistent, passive seismic sources via fiber-optic sensing \citep{liu2021retrievability, higueret2025trains, mukherjee2026anthropogenic}.

Bridge-radiated wavefields provide practical applications for civil infrastructure and hydrogeology. Foundation scour remains a primary driver of bridge collapse \citep{wardhana2003bridgefailures,prendergast2014scourreview}, operating through stiffness degradation of the supporting soil-pile system \citep{prendergast2013pilescourfrequency}. The high repeatability of wavefields near bridge piers could support future monitoring of localized foundation stiffness changes. In hydrogeology, the spatial $\mathrm{d}v/v$ results (Figure~\ref{fig:dvv}b) provide spatially distributed observations that may help constrain processes such as infiltration, pore-pressure changes, and vadose-zone wetting, complementing sparse well observations \citep{clements2018tracking,dangeard2021rivercorridor}. Nevertheless, our approach has limitations including lower sensitivity to the uppermost few meters of soil due to the modal frequency range, which higher-frequency vehicle-induced surface waves (used here up to $10\text{~Hz}$) can partly offset, as well as a reliance on routine traffic excitation for continuous passive illumination, whose diurnal and weekly volume cycles modulate hourly $\mathrm{d}v/v$. Selecting outward-propagating waves favors bridge-related illumination but does not fully remove source variability or other traffic-induced energy. The diurnal modulations therefore indicate that source effects must be evaluated alongside the flood-related medium response. Because the deck channels record structural vibrations directly, they provide a means to track source variability and assess its influence on the riverbank measurements. Overall, leveraging existing telecommunication dark fiber near bridge crossings offers a continuous, non-invasive platform for joint structural and near-surface hydrogeophysical monitoring.

\section{Conclusions}
\label{sec:conclusions}

This study demonstrates that traffic-excited bridge resonances recorded on dark fiber-optic cables provide continuous, energetic seismic illumination for near-surface imaging and hydrogeophysical monitoring. Functioning as a spatially fixed and highly repeatable source, the bridge fills the frequency gap between low-frequency microseisms and high-frequency traffic noise. Using seismic interferometry and surface wave inversion, we constrained the corridor $V_S$ structure to depths encompassing the interpreted bedrock transition. Wavefield repeatability supported velocity monitoring ($\mathrm{d}v/v$) using 1-hour correlation stacks during a major monsoon flood, while ballistic correlations converged within 5 minutes. The spatially distributed measurements revealed velocity reductions of up to $1\%$ associated with the flood, consistent with changes in riverbank saturation and pore pressure. Source-related modulation remains a limitation when interpreting the amplitude and spatial pattern of these changes. By converting existing telecommunication networks into continuous observatories, bridge-infrastructure seismology offers a scalable, non-invasive framework for tracking dynamic river-groundwater interactions, riverbank stability, and subsurface response to extreme weather.

\section{Data and Resources}
The DAS data used in this study are available upon request. Rainfall records were obtained from the Korea Meteorological Administration (\url{https://data.kma.go.kr}), and river-stage data from the Water Resources Management Information System (\url{http://www.wamis.go.kr}); both websites were last accessed in October 2026. The supplemental material includes examples of bridge- and vehicle-generated wavefields under contrasting traffic conditions and a comparison of bridge-source and vehicle-source $\mathrm{d}v/v$ results.

\section{Declaration of Competing Interests}
The authors declare that there are no competing interests with respect to this study.

\begin{ack}
H.L. and B.B. acknowledge support from the Stanford Earth Imaging Project (SEP) affiliate companies. H.L. also acknowledges support from the Jon Claerbout Fellowship. We thank Fu Yin for providing the ambient noise code. We also acknowledge support from the National Research Foundation of Korea (NRF) through a grant funded by the Korean government under award number RS-2024-00411986.
\end{ack}

\bibliography{references}

\end{document}